\documentstyle[12pt,psfig]{article}

\catcode`\@=11
\def\tripint#1{{d^d{\bf #1} \over{(2\pi)^d}}}

\def\simless{\raise.3ex\hbox{$<$\kern-.75em\lower 1ex\hbox{$\sim$}}}
\def\grsim{\raise.3ex\hbox{$>$\kern-.75em\lower 1ex\hbox{$\sim$}}}
\def\@maketitle{\newpage   
 \null
 \vspace*{-1\headsep}      
 \vspace*{-1\headheight}
 \vspace*{-24pt}
 \begin{flushright}{\absize
   { \preprintno} \\ \@date}
 \end{flushright}
 \vskip \headsep	   
 \vskip \headheight
 \begin{center}		   
   {\tisize\bf \@title \par}
   \vskip 2em
   {\ausize
     \begin{tabular}[t]{c}\@author
     \end{tabular}\par}
   \vskip 1ex
 \end{center}
 \par
 \vskip 2ex}

\newcommand{\preprintno}{}	 

\def\abstract{\if@twocolumn
\section*{Abstract}
\else \absize
\begin{center}
{\bf Abstract\vspace{0pt}}
\end{center}
\setskip
\quotation
\fi}
\def\endabstract{\if@twocolumn\else\endquotation\fi}

\def\section{
\setcounter{equation}{0} 	
\@startsection {section}{1}{\z@}{-3.5ex plus -1ex minus
 -.2ex}{2.3ex plus .2ex}{\Large\bf}}

\def\tisize{\Large}      
\def\ausize{\large}      
\def\absize{\normalsize} 
\def\setskip{ \setlength{\baselineskip}{4ex} } 

\def\starttext{
\setlength{\baselineskip}{ 17pt} 
\pagenumbering{arabic}
}

\let\<=\langle
\let\>=\rangle

\def\beq{\begin{equation}}
\def\eeq{\end{equation}}
\def\ba{\begin{array}}
\def\bea{\begin{eqnarray}}
\def\ea{\end{array}}
\def\eea{\end{eqnarray}}

\def\comment#1{ \hbox{[{\it Comment suppressed here.}\/]} }
\def\tr{\hbox{tr}\,}

\def\eqn#1{(\ref{#1})}  

\def\half{{\txt{1\over 2}}}

\begin{document}

\setcounter{footnote}{1}
\title{Self-Avoiding Walks with Writhe}

\author{J.~David Moroz and Randall D. Kamien\thanks{Corresponding author.
\tt kamien@dept.physics.upenn.edu}
        \\[1ex]
	\sl Department of Physics and Astronomy\\
        \sl University of Pennsylvania\\
        \sl Philadelphia, PA  19104\\[3ex]
}
\date{7 May 1997; modified 11 May 1997}

\renewcommand{\preprintno}{~~}

\begin{titlepage}

\maketitle

\def\thepage {}        

\begin{abstract}
We map self-avoiding random walks with a chemical potential for
writhe to the three-dimensional complex $O(N)$ Chern-Simons theory as
$N\rightarrow 0$.  We argue that at the Wilson-Fisher fixed point
which characterizes normal self-avoiding walks (with radius
of gyration exponent $\nu\approx 0.588$) a small chemical potential for writhe
is irrelevant and the Chern-Simons field does not modify the monomer-monomer
correlation function.  For a large chemical potential the polymer collapses.

\end{abstract}

\end{titlepage}
\starttext
\def\div{\nabla\!\cdot\!}
%
%
%
%
\input{epsf.sty}
%
%
%
%
%
%
%
%
%
%
%
%
%
%
%
%
%
%
%
%
%
%
%
%
%
%
%
%
%
%
%
%
%
%
%
%
%
%
%
%
%
%
%
%
%
%
%
%
%
%
%
%
%
%
%
%
%
%
%
%
%
%
%
%
%
%
%
%
%
%
%
%
%

%
%
%
%

%

%
%
%
%
%
%
\def\axowidth{0.5 }
\def\axoscale{1.0 }
\def\axoxoff{0 }
\def\axoyoff{0 }
\def\axoxo{0 }
\def\axoyo{0 }
\def\firstcall{1}
\def\Gluon(#1,#2)(#3,#4)#5#6{
%
%
\put(\axoxoff,\axoyoff){
}
}
\def\Photon(#1,#2)(#3,#4)#5#6{
%
%
\put(\axoxoff,\axoyoff){
}
}
\def\ZigZag(#1,#2)(#3,#4)#5#6{
%
%
\put(\axoxoff,\axoyoff){
}
}
\def\PhotonArc(#1,#2)(#3,#4,#5)#6#7{
%
%
\put(\axoxoff,\axoyoff){
}
}
\def\GlueArc(#1,#2)(#3,#4,#5)#6#7{
%
%
\put(\axoxoff,\axoyoff){
}
}
\def\ArrowArc(#1,#2)(#3,#4,#5){
%
%
\put(\axoxoff,\axoyoff){
}
}
\def\LongArrowArc(#1,#2)(#3,#4,#5){
%
%
\put(\axoxoff,\axoyoff){
}
}
\def\DashArrowArc(#1,#2)(#3,#4,#5)#6{
%
%
\put(\axoxoff,\axoyoff){
}
}
\def\ArrowArcn(#1,#2)(#3,#4,#5){
%
%
\put(\axoxoff,\axoyoff){
}
}
\def\LongArrowArcn(#1,#2)(#3,#4,#5){
%
%
\put(\axoxoff,\axoyoff){
}
}
\def\DashArrowArcn(#1,#2)(#3,#4,#5)#6{
%
%
\put(\axoxoff,\axoyoff){
}
}
\def\ArrowLine(#1,#2)(#3,#4){
%
%
\put(\axoxoff,\axoyoff){
}
}
\def\LongArrow(#1,#2)(#3,#4){
%
%
\put(\axoxoff,\axoyoff){
}
}
\def\DashArrowLine(#1,#2)(#3,#4)#5{
%
%
\put(\axoxoff,\axoyoff){
}
}
\def\Line(#1,#2)(#3,#4){
%
%
\put(\axoxoff,\axoyoff){
}
}
\def\DashLine(#1,#2)(#3,#4)#5{
%
%
\put(\axoxoff,\axoyoff){
}
}
\def\CArc(#1,#2)(#3,#4,#5){
%
%
\put(\axoxoff,\axoyoff){
}
}
\def\DashCArc(#1,#2)(#3,#4,#5)#6{
%
%
\put(\axoxoff,\axoyoff){
}
}
\def\Vertex(#1,#2)#3{
%
%
\put(\axoxoff,\axoyoff){
}
}
\def\Text(#1,#2)[#3]#4{
%
%
\dimen0=\axoxoff \unitlength
\dimen1=\axoyoff \unitlength
\advance\dimen0 by #1 \unitlength
\advance\dimen1 by #2 \unitlength
\makeatletter
\@killglue\raise\dimen1\hbox to\z@{\kern\dimen0 \makebox(0,0)[#3]{#4}\hss}
\ignorespaces
\makeatother
}
\def\BCirc(#1,#2)#3{
%
%
\put(\axoxoff,\axoyoff){
}
}
\def\GCirc(#1,#2)#3#4{
%
%
\put(\axoxoff,\axoyoff){
}
}
\def\EBox(#1,#2)(#3,#4){
%
%
\put(\axoxoff,\axoyoff){
}
}
\def\BBox(#1,#2)(#3,#4){
%
%
\put(\axoxoff,\axoyoff){
}
}
\def\GBox(#1,#2)(#3,#4)#5{
%
%
\put(\axoxoff,\axoyoff){
}
}
\def\Boxc(#1,#2)(#3,#4){
%
%
\put(\axoxoff,\axoyoff){
}
}
\def\BBoxc(#1,#2)(#3,#4){
%
%
\put(\axoxoff,\axoyoff){
}
}
\def\GBoxc(#1,#2)(#3,#4)#5{
%
%
\put(\axoxoff,\axoyoff){
}
}
\def\SetWidth#1{\def\axowidth{#1 }}
\def\SetScale#1{\def\axoscale{#1 }}
\def\SetOffset(#1,#2){\def\axoxoff{#1 } \def\axoyoff{#2 }}
\def\SetScaledOffset(#1,#2){\def\axoxo{#1 } \def\axoyo{#2 }}
\def\pfont{Times-Roman }
\def\fsize{10 }
\def\SetPFont#1#2{\def\pfont{#1 } \def\fsize{#2 }}
%
%
\makeatletter
\def\fmode{4 }
\def\@l@{l} \def\@r@{r} \def\@t@{t} \def\@b@{b}
\def\mymodetest#1{\ifx#1\end \let\next=\relax \else {
\if#1\@r@\global\def\fmodeh{-3 }\fi
\if#1\@l@\global\def\fmodeh{3 }\fi
\if#1\@b@\global\def\fmodev{-1 }\fi
\if#1\@t@\global\def\fmodev{1 }\fi
} \let\next=\mymodetest\fi \next}
\makeatother
\def\PText(#1,#2)(#3)[#4]#5{
%
%
\def\fmodev{0 }
\def\fmodeh{0 }
\mymodetest#4\end
\put(\axoxoff,\axoyoff){\makebox(0,0)[]{\special{"/\pfont findfont \fsize
 scalefont setfont #1 \axoxo add #2 \axoyo add #3
\fmode \fmodev add \fmodeh add \fsize (#5) \axoscale ptext }}}
}
\def\GOval(#1,#2)(#3,#4)(#5)#6{
%
%
\put(\axoxoff,\axoyoff){
}
}
\def\Oval(#1,#2)(#3,#4)(#5){
%
%
\put(\axoxoff,\axoyoff){
}
}
\let\eind=]
\def\DashCurve#1#2{\put(\axoxoff,\axoyoff){
}}
\def\Curve#1{\put(\axoxoff,\axoyoff){
}}
\def\kromme(#1,#2)#3{#1 \axoxo add #2 \axoyo add \ifx #3\eind\else
\expandafter\kromme\fi#3}
\def\LogAxis(#1,#2)(#3,#4)(#5,#6,#7,#8){
%
%
\put(\axoxoff,\axoyoff){
}
}
\def\LinAxis(#1,#2)(#3,#4)(#5,#6,#7,#8,#9){
%
%
\put(\axoxoff,\axoyoff){
}
}
\input rotate.tex
\makeatletter
\def\rText(#1,#2)[#3][#4]#5{
%
%
\ifnum\firstcall=1\global\def\firstcall{0}\rText(-10000,#2)[#3][]{#5}\fi
\dimen2=\axoxoff \unitlength
\dimen3=\axoyoff \unitlength
\advance\dimen2 by #1 \unitlength
\advance\dimen3 by #2 \unitlength
\@killglue\raise\dimen3\hbox to \z@{\kern\dimen2
\makebox(0,0)[#3]{
\ifx#4l{\setbox3=\hbox{#5}\rotl{3}}\else{
\ifx#4r{\setbox3=\hbox{#5}\rotr{3}}\else{
\ifx#4u{\setbox3=\hbox{#5}\rotu{3}}\else{#5}\fi}\fi}\fi}\hss}
\ignorespaces
}
\makeatother
\def\BText(#1,#2)#3{
%
%
\put(\axoxoff,\axoyoff){
}
}
\def\GText(#1,#2)#3#4{
%
%
\put(\axoxoff,\axoyoff){
}
}
\def\B2Text(#1,#2)#3#4{
%
%
\put(\axoxoff,\axoyoff){
}
}
\def\G2Text(#1,#2)#3#4#5{
%
%
\put(\axoxoff,\axoyoff){
}
}
\catcode`\@=12
\def\Lk{\hbox{\sl Lk}}
\def\Wr{\hbox{\sl Wr}}
\def\Tw{\hbox{\sl Tw}}
\def\cross{\!\times\!}
\def\inner{\!\cdot\!}
\def\curl{\nabla\!\times\!}
\def\mod{\hbox{\rm mod}~}
\def\pt{\partial}
\def\lt{\left}
\def\rt{\right}
\def\perpp{{\scriptscriptstyle\perp}}
\def\kb{k_{\scriptscriptstyle\rm B}}
\def\tiny{\scriptscriptstyle\rm}
\def\der#1{{d{#1}(\ell)\over d\ell}}
\def\half{{1\over 2}}
\def\nml{\hbox{$\cal N$}}
\def\ham{\hbox{$\cal H$}}
\def\kbT{k_{\scriptscriptstyle\rm B}T}
\def\TT{\hbox{$\widetilde T$}}
\def\bo#1{{\cal O}(#1)}
\def\th{{\bf\hat t}}
\def\dnb{\delta\vec n}
\def\ofx{({\vec x})}
\def\ofxp{({\vec x}')}
\def\ofxt{}
\def\hef{$^4${\rm He}}
\def\het{$^3${\rm He}}
\def\lb{\hbox{$\bar\lambda$}}
\def\rsq{\hbox{$\overline{\langle\,\left( r(t) - r(0) \right)^2\,\rangle}$}}
\def\free{\hbox{$\cal F$}}
\def\bold#1{\setbox0=\hbox{$#1$}%
     \kern-.010em\copy0\kern-\wd0
     \kern.025em\copy0\kern-\wd0
     \kern-.020em\raise.0200em\box0 }

\def\grad{\bold{\nabla}}
\def\brac#1#2{\{{#1},{#2}\}}
\def\thth#1{\Theta\left[z-s_{#1}\right]\Theta\left[e_{#1} -z\right]}

\section{Introduction and Summary}
\indent Aside from their mathematical interest, self-avoiding random walks are
good models for polymers which are much longer than their persistence
length.  In this paper, we study how these self-avoiding random walk
models can be adapted to study the effects of a topological constraint
on the polymer.  Recent progress in experimental technique has, for
the first time, allowed the direct, real-space observation of polymer
conformations
\cite{Kas}.  These studies, as well as experiments in which polymers
have been subjected to torsional constraints \cite{Strick}, have
focussed primarily on long biopolymers.  While
self-avoidance is important in long biomolecules such as actin
or DNA, these molecules, in addition, have a {\sl twist} rigidity.  If
the twist were decoupled from the conformational changes of the
polymer then twist rigidity would be unimportant.  However, for a
closed polymer, such as a plasmid DNA loop, there is a topological
invariant, namely the linking number of the two DNA backbones.
Fuller's \cite{Fuller}\ now ubiquitous
\cite{Marko,Rudnick1,Rudnick2,Julicher}\ relation $\Lk=\Tw+\Wr$ provides
a connection between the topological linking number $\Lk$, the total
amount of twist in the polymer $\Tw$ and the writhe of the curve
$\Wr$---a scalar which captures some geometrical information about the
conformational deformations of the loop.
The way that topological constraints can affect the conformational
statistics of polymers is now accessible to experimental
investigation.  Studies of DNA force-extension curves with constrained
$\Lk$ have discovered a coupling between an excess in linking number
and the backbone configuration \cite{Strick}.  The question arises:
could there be a new universality class describing the scaling regime
of topologically constrained polymers?
\par
There is currently a wealth of information on the statistics of
self-avoiding random walks.  In particular, a most remarkable mapping
by de Gennes \cite{deGennes}\ between the critical behavior of the
scalar $O(N)$ model as $N\rightarrow 0$ and the behavior of
self-avoiding walks has allowed, via any number of schemes, a method
of calculating the radius of gyration exponent $\nu$ which measures
how the radius of gyration scales with the polymer length: $R_G\sim
L^\nu$.  It is interesting then to consider the effect of the
topological constraint on the radius of gyration of the closed polymer
loop.  The topological constraint can be added to the standard
analysis via a gauge field, as noted by Brereton and Shah \cite{BS}.
\par
As we will argue, polymers with
a small chemical potential for $\Lk$ fall into the same universality
class as the usual self-avoiding random walk.  The experimentally
observed coupling arises from the non-universal details of the polymer and
not from any general long-wavelength properties.  A chemical potential
for $\Lk$ only becomes noticeable when it exceeds a certain
threshold, at which point the chain collapses \cite{Marko}.
\par
So far we have focussed our attention on a chemical potential for
$\Lk$.  In what follows, we turn our attention to the writhe of a
curve.  The introduction of a Chern-Simons gauge field allows us to
naturally and easily incorporate a chemical potential for writhe
\cite{Witten}.  In section 2 we will show that
a chemical potential for writhe is equivalent to a chemical potential
for link.  What is more, our model will only have one backbone to keep
track of.
\par
In section 3 we map the statistics of self-avoiding walks with a
chemical potential for their writhe onto the $N\rightarrow 0$ complex
scalar Chern-Simons theory.  We then argue in section 4
that at the self-avoiding
fixed point controlled by $\nu\approx 0.588$ \cite{Sokal} the effect of writhe
is
irrelevant for a small chemical potential.  From this we conclude that
there is no coupling between the conformations of a polymer and its
writhe in the scaling regime.  Finally, in section 5 we calculate moments of
the writhe and find that the field-theoretic analysis disagrees
with some rigorous results \cite{Orlandini1994,Rensburg1993}.
We understand this to be a
consequence of the writhe ``hiding'' from the long wavelength field
theory in persistence length size, plectonemic regions separated from each
other.

\section{Integrating Out the Twist}

\par The theory we develop in section 3 incorporates a chemical
potential for the writhe of a closed polymer loop.  While it would be
difficult, experimentally, to add such a chemical potential, a chemical
potential for linking number could easily be arranged.  For example,
DNA plasmids in the presence of topoisomerase and ATP will tend to
have some non-zero excess linking number due to the bias of the added
enzyme.  We show in this section that a chemical potential for writhe
is equivalent to a chemical potential for the linking number.
\par
For simplicity we assume that the long-wavelength behavior of the
polymer is adequately described by a simple continuum model.
The free energy for this model is determined by a bending stiffness
$\kappa$ and a twist rigidity $C$\cite{Marko}:
\beq
F[{\bf R},\Omega]={1 \over 2}\int ds\left\{\kappa\left({{d^2 {\bf
R}}\over{ds^2}}\right)^2
+ C \Omega^2 \right\} ,
\eeq
where ${\bf R}(s)$ describes the conformation of the chain backbone,
$\Omega (s)$ is the twist degree of freedom and $s$ is the curves arclength.
The partition
function for this model with a chemical potential $g^2$ for the linking
number is then given by
\beq
Z=\int[d{\bf R}][d\Omega]\,\exp\left\{-F[{\bf R},\Omega]
-{g^2} \left(\Wr[{\bf R}]+{1\over 2\pi}\int\Omega ds\right)\right\} ,
\eeq
where the energy is measured in units of $k_{\tiny B}T$.  We have used
Fuller's formula $\Lk=\Tw+\Wr$ explicitly to divide the linking number
into the integral of the twist density $\Omega$ and the writhe,
where the writhe of the curve is
\beq\label{ewri}
\Wr[{\bf R}]={1\over 4\pi}
\int{ds\,\int{ds'\,{\left({{{d{\bf R}(s)} \over {ds}} \cross {{d{\bf R}(s')}
\over {ds'}}} \right)
\inner {\left[{\bf R}(s) - {\bf R}(s')\right] \over \left\vert {\bf R}(s)-{\bf
R}(s')\right\vert^3}}}}.
\eeq
The twist degree of freedom can be integrated out
to yield a new effective partition function with a chemical potential for
writhe:
\beq\label{Zeff}
Z_{\rm eff}=\int [d{\bf R}] \exp\left\{-{\kappa\over 2}\int
\left({{d^2{\bf R}} \over{ds^2}}\right)^2ds -{g^2} \Wr[{\bf R}]
\right\}.
\eeq
Unlike the twist, the writhe is non-local and can not be integrated out so
easily.
\par
In other models it may be somewhat more cumbersome to integrate out the
twist degree of freedom.  As an example we can consider a ribbon-like
model.  Once again we will parameterize the backbone by ${\bf R}(s)$,
but we will use a unit vector ${\bf u}$(s) to keep track of the twist.
We take this vector to be perpendicular to the backbone tangent, for
instance, in DNA ${\bf u}$ could point from the central backbone to
one of the two sugar-phosphate strands.  In terms of this parameterization, the
twist density is
\beq
\Omega={{d{\bf R}}\over{ds}} \inner \left({{d{\bf u}}\over{ds}} \cross {\bf
u}\right) .
\eeq
The partition function for this modified model is\\
\vbox{\bea
Z&=&\int\left[d{\bf R}\right]\left[d\Omega\right]\left[d{\bf u}\right]
\exp\left\{-F\left[{\bf R},\Omega\right]-{g^2} \left(\Wr[{\bf
R}] +
{1\over 2\pi}\int\Omega\,ds \right)\right\} \nonumber \\
&& \qquad\times\delta\left[\Omega-{{d{\bf R}}\over{ds}} \inner \left({{d{\bf
u}}\over{ds}}
\cross {\bf u}\right)\right]\,\delta[{\bf u}^2-1]\,\delta\left[{\bf u}
\inner {{d{\bf R}} \over {ds}}\right] .
\eea}
Upon integrating out ${\bf u}$,\\
\vbox{\bea
Z&=&\int[d{\bf R}][d\Omega]{\det}^{-1}\left[{\delta
\left[\Omega-{{d{\bf R}}\over{ds}} \inner \left({{d{\bf u}}\over{ds}}
\cross {\bf u}\right),{\bf u}^2-1,{\bf u}\inner{{d{\bf R}} \over {ds}}\right]
} \over {\delta{\bf u}}\right] \nonumber \\
&&\qquad\times\exp\left\{-F[{\bf R},\Omega]-{{g^2}
}\left(\Wr[{\bf R}]
+{1\over 2\pi}\int ds\,\Omega\right)\right\}.
\eea}
The determinant is an isotropic functional of the backbone tangent.
This functional can be expanded in powers of $d{\bf R}/ds$.  Since at
distances large compared to the persistence length any local isotropic
model flows to the random walk fixed point we will, again, only consider the
effective partition function given in \eqn{Zeff}.

\section{Mapping to $N\rightarrow 0$ Scalar Chern-Simons Theory}\label{mapping}

\par In this section, we map the $N \rightarrow 0$, $O(N)$ complex scalar
field theory with a Chern-Simons gauge field to a self-avoiding random
walk with a chemical potential for writhe.  The Chern-Simons term is
introduced because after functionally-integrating it out
we obtain an expression for the writhe of the
self-avoiding random walk described by the scalar theory
\cite{BS,Witten}.  A similar procedure to this one was used to study the
area of walks in two-dimensions by Cardy \cite{Cardy} and
the linking number of a pair of walks where one walk fills
space \cite{BS}, though in both those cases
the gauge field had the standard Maxwell potential $F_{\mu\nu}^2$.
\par
We start by considering the Wilson loop along a closed path
$\Gamma$ for a Chern-Simons gauge theory with coupling $g$.  Namely
\bea\label{ewilop}
C[\Gamma]&\equiv&\left\langle\,\exp\left\{i\oint_\Gamma dR_\mu
A^\mu\right\}
\right\rangle\\
&=& \int [d{\bf A}]\,\exp\left\{i\int ds {dR_\mu\over ds}
\inner A^\mu \left({\bf R}(s)\right)\right\}
\exp\left\{ -{1\over 4g^2}\int d^3\!x\, {\bf A}\inner\curl{\bf A}\right\}\,
\exp\left\{-S_{\rm gf}\right\}, \nonumber
\eea
where we have added the gauge fixing term,
\beq
S_{\rm gf} = {1 \over {2\Delta}}\int d^3\!x\, (\nabla \inner {\bf A})^2 .
\eeq
Though the Wilson loop is gauge-invariant, in practice we must select
a gauge to compute it.  We choose the Landau gauge ($\Delta
\rightarrow 0$) so that the Chern-Simons propagator takes on
a simple form:
\beq\label{gaugeprop}
\langle{A_i(q) A_k(-q)}\rangle=2g^2{{i \epsilon_{ijk} q_j} \over q^2} +\Delta
{{q_i q_k} \over q^4} \stackrel{\Delta \rightarrow 0}
{\longrightarrow} 2g^2{{i \epsilon_{ijk} q_j} \over q^2}.
\eeq
As we shall show, the antisymmetry of the propagator
in this gauge simplifies the renormalization group analysis by
eliminating a number of Feynman graphs.
\par
Writing
\beq\label{etrick}
A^\mu\left({\bf R}(s)\right)=
\int d^3\!x\,\delta^3\left[{\bf x}-{\bf R}(s)\right]
A^\mu({\bf x})
\eeq
we can perform the functional integration in \eqn{ewilop}\ to find
\beq\label{eintwil}
\left\langle\,\exp\left\{ i\oint_\Gamma dR_\mu
A^\mu\right\}\,\right\rangle
= \exp\left\{ -{g^2} \Wr[\Gamma]\right\} .
\eeq
The writhe of any closed path is therefore determined by the line
integral of the Chern-Simons gauge field.
\par
To include self-avoidance, we employ a generalization of de Gennes'
mapping of self-avoiding random walks onto the $N\rightarrow 0$, $O(N)$
real scalar field theory \cite{deGennes}.
It is useful instead to consider the $N\rightarrow 0$, $O(N)$ {\sl
complex} scalar field theory which is equivalent\cite{Miller}.
Recall that the mapping proceeds by considering the high temperature
series for the $N$ component complex Ising model with Hamiltonian
\beq\label{eheiham}
H = \sum_{\langle i,j\rangle} {\bf s}_i^\dagger\inner {\bf s}_j + {\rm
c.c}
\eeq
where $\langle i,j\rangle$ denotes nearest neighbors and ${\bf s}_j$
is an $N$-component complex vector field with ${\bf s}_i^\dagger\inner{\bf
s}_i = N$.  The discrete version of the Wilson line for a path
from ${\bf 0}$ to ${\bf x}$ is
\beq\label{ediswil}C[\Gamma] = \left\langle
\exp\{i\sum_{\Gamma} A_{ij}\}\right\rangle
\eeq
where $A_{ij}$ is a field defined on the link connecting sites $i$ and $j$
and the sum is along the path $\Gamma$ which connects ${\bf 0}$ and ${\bf
x}$.  In order to weight each path by its writhe, we can include in the
series expansion
the corresponding Wilson line and integrate out ${\bf A}$ to
give each path a chemical potential $g^2$ for writhe.  Incorporation
of this phase is straightforward---we replace the Hamiltonian in \eqn{eheiham}\
with a modified one:
\beq\label{eguaham}
H =\sum_{\langle i,j\rangle} {\bf s}_i^\dagger\inner{\bf s}_j U_{ij} + {\rm
c.c.},
\eeq
where $U_{ij}= \exp\{iA_{ij}\}$.  We recognize this Hamiltonian as that for a
lattice gauge theory \cite{Kogut}, where the local gauge symmetry is
\bea\label{egua}
U_{ij}\,&\rightarrow&U_{ij} e^{i\left(\theta_i-\theta_j\right)}\nonumber\\
{\bf s}_i\,&\rightarrow& {\bf s}_ie^{i\theta_i}
\eea
where $\theta_i$ is arbitrary.  In the continuum limit the
transformation of $U_{ij}$ becomes ${\bf A}\rightarrow {\bf A} +
\partial\theta$, the usual gauge transformation.  Taking the
continuum limit \cite{Kogut} of \eqn{egua}\ and including the
Chern-Simons term leads us to the $O(N)$ complex scalar Chern-Simons
action
\beq\label{euncs}
S=\int d^3\!x\,\left(e \left\vert\left(\partial-i{\bf
 A}\right)\vec\psi\right\vert^2 -\mu\vert\vec\psi\vert^2 +
 u\vert\vec\psi\vert^4 + {1\over 4g^2} {\bf A}\inner\curl{\bf A}\right),
\eeq
where $\vec\psi$ is an $N$-component, complex scalar field, $u$ is the
monomer-monomer
repulsion and $\mu$ is the chemical potential to add a monomer or step.
Note that, though the scalar field theory has a $U(N)$ symmetry, we are
only gauging the diagonal $U(1)$.
There is no quantization condition on $g$ as is
typical in non-Abelian Chern-Simons theories \cite{jackiw} or in the
presence of magnetic monopoles \cite{quanti1,quanti2}, which are not part of
this theory.
\par
Recalling the mapping \cite{deGennes,Miller}, the correlation function is
\beq\label{ewilli}
\widetilde G_\Gamma({\bf x})=\lim_{N\rightarrow 0}
{1\over N}\left\langle\,\vec\psi^\dagger({\bf x})\inner\vec\psi({\bf 0})
\,\right\rangle = \sum_L \Upsilon({\bf x};L)e^{\mu L}
\eeq
where $\Upsilon({\bf x};L)$ is the sum over all paths of length $L$
that connect ${\bf 0}$ to ${\bf x}$ weighted by
$\exp\{-g^2\Wr\}$ and $(\cdot)$ is the inner product defined on
the $N$-component internal space.  This correlation function is not
gauge invariant: to make it gauge invariant we could add a Wilson line
to the correlation function.  The arbitrariness of the path we choose
to connect $\bf 0$ to $\bf x$ will lead to an arbitrary value of the
writhe since the expression in \eqn{ewri}\ is manifestly not
additive---it depends on the whole curve.
\par
For a closed polymer loop, however, we can consider the correlation
function
\bea\label{eclosed}
G^C_\Gamma({\bf x})&&\equiv
{1\over N^2}\sum_{\alpha\beta}\left\langle\,s^\alpha_{\bf
0}\left(s^\beta_{\bf 0}\right)^*\left(s^\alpha_{\bf x}
\right)^*s^\beta_{\bf x}
\,\right\rangle\nonumber\\
&&\rightarrow {1\over N^2}\left\langle\,\vec
\psi^\dagger({\bf
x})\inner\vec\psi({\bf 0})\vec\psi^\dagger({\bf 0})
\inner\vec\psi({\bf x})\,\right\rangle
\eea
where $\alpha$ and $\beta$ are the indices of the $N$ component field and
$(\cdot)$ is the inner product defined on the $N$-component internal space.
This correlation function is gauge invariant and, in the $N\rightarrow
0$ limit may be written as
\beq\label{ewillii}
G^C({\bf x}) = \sum_L \Upsilon^C({\bf x};L)e^{\mu L},
\eeq
where now $\Upsilon^C({\bf x};L)$ is the sum
over all closed loops of length $L$ which pass
through $\bf 0$ and $\bf x$ weighted by $\exp\{-g^2\Wr\}$.
Upon normalizing $\Upsilon^C({\bf x};L)$
for fixed $L$ we may find the probability of a closed, self-avoiding
walk passing through $\bf 0$ and $\bf x$ with a chemical potential
$g^2$ for its writhe.  For a monomer chemical potential $\mu$
close to its critical value $\mu_c$, $G^C({\bf x},\mu)$ will scale as
$[\vert{\bf x}\vert(\mu-\mu_c)^\nu]$, where $\nu$ is the usual
correlation length exponent of critical phenomena.  Inverting the sum
in \eqn{ewilli} to find $\Upsilon^C_L({\bf x})$ will lead to a scaling
form for the probability
\beq\label{eprobb}
P^C({\bf x};L) \propto \Xi\left({\vert{\bf x}\vert\over L^\nu}\right)
\eeq
and so the radius of gyration of the closed self-avoiding walk will
still scale as $L^\nu$.  In the next section we will argue that at the
self-avoiding walk fixed point ({\sl a.k.a} the Wilson-Fisher fixed
point) the Chern-Simons gauge field is {\sl irrelevant}, and thus the
presence of writhe does not change the radius of gyration of a closed
polymer loop for $g\simless 1$.

\section{Renormalization Group Results}

\par In this section we perform a renormalization group analysis of the
complex $O(N)$ spin model coupled to a Chern-Simons gauge field in the
$N \rightarrow 0$ limit.  We will show that the presence of the
Chern-Simons term does not change the long-distance behavior of the four
point correlation function in \eqn{eclosed}.
\par
Our analysis is based on a perturbation expansion around the
Wilson-Fisher fixed point of the ungauged model.  At this fixed point
we have the following symbolic recursion relations which define the critical
behavior for self-avoiding random walks:
\bea\label{recursion}
{de} \over {dl} &=& e~\left(-\eta+f_e(e,u)\right) , \nonumber \\
{d\mu} \over {dl} &=& \mu~f_\mu(e,u) , \nonumber \\
{du} \over {dl} &=& u~f_u(e,u) .
\eea
The functions $f_{\rm x}$ can be computed, for instance, by a
$d=4-\epsilon$ expansion \cite{WilsonFisher}.  The scaling relation
given in \eqn{recursion} for the coefficient $e$ defines an anomalous
dimension for the scalar fields: if this coefficient is chosen not to rescale,
then its exponent is absorbed into the $\psi$ and
$\psi^\dagger$ fields.  This implies that the scalar propagator
no longer scales with its na\"\i ve dimension: at the
Wilson-Fisher fixed point the scalar propagator is proportional to
$k^{-2+\eta}$.  This makes the infrared power counting at the Wilson-Fisher
fixed point softer than that at the Gaussian fixed point where
$\langle\psi^\dagger\psi\rangle \sim k^{-2}$.  As we
shall see, this difference will prove important by keeping the
corrections to \eqn{recursion} from diverging in the infrared.
\par
In what follows we will consider how the addition of the Chern-Simons
gauge field modifies these recursion relations for the self-avoiding
walk field theory in three dimensions.

\subsection{Non-Renormalization of the Gauge Field}
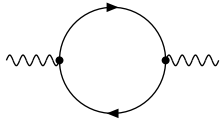
\begin{figure}[t]
\begin{center}
\begin{picture}(100,50)(0,0)
\ArrowArcn(50,25)(20,180,0)
\ArrowArcn(50,25)(20,360,180)
\Vertex(30,25){1.5}
\Vertex(70,25){1.5}
\Photon(10,25)(30,25){2}{4}
\Photon(70,25)(90,25){2}{4}
\end{picture}
\end{center}
\caption{\label{loopzero} A correction to the gauge propagator with a closed
scalar loop which vanishes in the $N \rightarrow 0$ limit.}
\end{figure}

We first show that the gauge propagator is not renormalized.  This
may not be surprising since the gauge field can only couple to closed
scalar loops.  These scalar loops generate diagrams proportional to
$N$ which vanish in the $N \rightarrow 0$ limit (fig.~\ref{loopzero}).
To prove that the gauge field does not get renormalized, we introduce
an auxiliary Hubbard-Stratonovich field $\chi$ through a new term
that we will add to the
action in \eqn{euncs}:
\beq
S_\chi = {1 \over 2}\int {d^3\!x~\left(\chi + i \sqrt{2u}
\vert\psi\vert^2 \right)^2} .
\eeq
Since the $\chi$ field enters quadratically into the action, the new
partition function will simply be proportional to the old one and the
equations of motion for $\psi$, $\psi^\dagger$, and ${\bf A}$ will be
left unchanged:\\
\vbox{\bea
Z&=&\int[d{\bf A}][d\chi]\left[d\vec\psi\right]\left[d\vec\psi^\dagger\right]
\exp
\left\{-\int{d^3\!x\,\left[e\left\vert\left(\partial-i{\bf
A}\right)\vec\psi\right\vert^2
-\left(\mu-i \chi \sqrt{2
u}\right)\left\vert\vec\psi\right\vert^2\right]}\right\} \nonumber \\
&&\qquad\times\exp\left\{-\int{d^3\!x\,\left({1 \over 4g^2}{\bf A} \inner \curl
{\bf A} +
{1 \over {2\Delta}} ({\bf \nabla} \inner {\bf A})^2 + {\chi^2 \over
2}\right)}\right\} .
\eea}
Since the scalar field now appears quadratically, we can integrate it out,
leaving a determinant:\\
\vbox{\bea\label{detZ}
Z&=&\int[d{\bf A}][d\chi]\,{{\det}^{-N}\left[e\left(\partial-i{\bf
A}\right)^2-\left(\mu-i \chi \sqrt{2 u}\right)\right]} \nonumber \\
&&\qquad\times\exp\left\{-\int {d^3\!x\,\left({1 \over 4g^2} {\bf A} \inner
\curl {\bf A} +
{1 \over {2\Delta}} ({\bf \nabla} \inner {\bf A})^2 + {\chi^2 \over
2}\right)}\right\} .
\eea}
In the $N \rightarrow 0$ limit, the determinant does not contribute to
the ${\bf A}-\chi$ partition function.  The Chern-Simons field ${\bf
A}$ is therefore a free field which keeps its na\"\i ve
dimension in the renormalization process. Moreover gauge invariance implies
that ${\bf A}$ must
scale as $\partial$ and so, in three dimensions, $g^2$ does not rescale:
\beq
{{dg}\over{dl}}=0 .
\eeq
Though the gauge coupling does not get renormalized, its presence may still
renormalize the couplings $e$, $\mu$, and $u$.  We will explore this
possibility
in the next subsection.
\par
\par
Superficially it might appear that the same argument presented above
for the ${\bf A}$ field could be applied to the scalar field $\psi$.
This would lead to the incorrect conclusion that $\psi$ also keeps its
na\"\i ve dimension upon renormalization.  Of course the $\psi$ field
does get renormalized -- it acquires an anomalous dimension $\eta$ as
indicated in \eqn{recursion}.  The argument put forward above cannot
be applied to the scalar field as this field's couplings are already
${\cal{O}}(N)$.  The determinant that we were previously able to
disregard generates just such terms, as can be seen from an expansion
of $\det^N[M]\equiv \exp(N \tr\ln M)$, where $M$ is the
operator in \eqn{detZ}.

\subsection{Perturbation Around the Wilson-Fisher Fixed Point}
\par
We start our renormalization group analysis with the recursion
relations for $e$, $\mu$, and $u$ at the Wilson-Fisher fixed point of
the ungauged $O(N)$ complex scalar theory in the $N \rightarrow 0$
limit \eqn{recursion}.  We will perform a one-loop perturbative expansion
in the gauge field propagator.
\par
\begin{figure}
\begin{center}
\begin{picture}(100,100)(0,0)
\ArrowLine(10,80)(43.16,80)
\ArrowLine(56.84,80)(90,80)
\ArrowArc(50,61.21)(20,270,70)
\ArrowArc(50,61.21)(20,110,270)
\Photon(50,41.21)(50,10){2}{6}
\DashLine(43.16,80)(56.84,80){1.5}
\Vertex(50,41.21){1.5}
\Text(0,100)[tl]{(a)}
\end{picture}
\begin{picture}(100,100)(0,0)
\ArrowLine(10,80)(43.16,80)
\ArrowLine(56.84,80)(90,80)
\ArrowArc(50,61.21)(20,270,70)
\ArrowArc(50,61.21)(20,110,270)
\Photon(50,41.21)(20,10){2}{7}
\Photon(50,41.21)(80,10){2}{7}
\DashLine(43.16,80)(56.84,80){1.5}
\Vertex(50,41.21){1.5}
\Text(0,100)[tl]{(b)}
\end{picture}
\begin{picture}(100,100)(0,0)
\ArrowLine(10,80)(43.16,80)
\ArrowLine(56.84,80)(90,80)
\ArrowArc(50,61.21)(20,315,70)
\ArrowArc(50,61.21)(20,110,225)
\ArrowArc(50,61.21)(20,225,315)
\Photon(35.86,47.07)(20,10){2}{7}
\Photon(64.14,47.07)(80,10){2}{7}
\DashLine(43.16,80)(56.84,80){1.5}
\Vertex(35.86,47.07){1.5}
\Vertex(64.14,47.07){1.5}
\Text(0,100)[tl]{(c)}
\end{picture}
\end{center}
\caption{\label{scalarloop} Corrections to the cubic and quartic
gauge vertices.}
\end{figure}
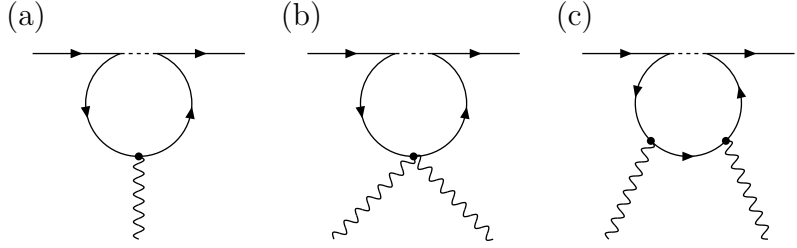
At zeroth order in our perturbation analysis there
is no gauge propagator.  The lack of a dynamical gauge field severely
limits the number of graphs that can modify the renormalization group
equations: to one loop, there are only three of them
(fig.~\ref{scalarloop}).  The first of these
(fig.~\ref{scalarloop}(a)) could contribute a correction to the single
gauge field vertex:
\beq\label{gcoupling}
\Gamma^{(0)}_{A\psi^\dagger\psi}=
\int \tripint{k_1} \tripint{k_2}\tripint{k_3}\,\delta^d({\bf k_1}+{\bf
k_2}+{\bf k_3})
A_i ({\bf k_3}) \vec\psi^\dagger({\bf k_2})\inner\vec\psi({\bf k_1}) ({\bf
k_1}-{\bf k_2})_i .
\eeq
The second and third graphs (figs.~\ref{scalarloop}(b) and
\ref{scalarloop}(c)) could modify the quartic coupling,
\vbox{\bea\label{g2coupling}
\Gamma^{(0)}_{AA\psi^\dagger\psi}&=&\int
\tripint{k_1}\tripint{k_2}\tripint{k_3}\tripint{k_4}\,
\delta^d({\bf k_1}+{\bf k_2}+{\bf k_3}+{\bf k_4}) \nonumber \\
&&\qquad \times A_i({\bf k_3}) A_i({\bf k_4})
\vec\psi^\dagger ({\bf k_2})\inner
\vec\psi({\bf k_1})  .
\eea}
\begin{figure}[b]
\begin{center}
\begin{picture}(100,50)(0,0)
\ArrowLine(10,20)(30,20)
\ArrowLine(30,20)(70,20)
\ArrowLine(70,20)(90,20)
\PhotonArc(50,20)(20,0,180){2}{8}
\Vertex(30,20){1.5}
\Vertex(70,20){1.5}
\end{picture}
\end{center}
\caption{\label{prop} Wavefunction renormalization of the $\psi$ field.}
\end{figure}
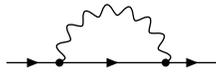
In a gauge invariant regularization scheme such as dimensional regularization,
fig.~\ref{scalarloop}(a) is identically
zero.  By gauge invariance, the other two graphs must then cancel.
Therefore to one loop order, the presence of a non-dynamical gauge
field does not modify the recursion relations given in \eqn{recursion}
and so leaves the critical behavior of our model unchanged.  This confirms that
$\bf A$ scales as $\partial$ when $\bf A$ is not dynamical.
\begin{figure}[b]
\begin{center}
\begin{picture}(100,100)(0,0)
\ArrowLine(10,50)(30,50)
\ArrowLine(30,50)(50,50)
\ArrowLine(50,50)(70,50)
\ArrowLine(70,50)(90,50)
\PhotonArc(50,50)(20,0,180){2}{8}
\Photon(50,50)(50,10){2}{6}
\Vertex(30,50){1.5}
\Vertex(50,50){1.5}
\Vertex(70,50){1.5}
\Text(0,100)[lt]{(a)}
\end{picture}
\begin{picture}(100,100)(0,0)
\ArrowLine(10,50)(30,50)
\ArrowLine(30,50)(70,50)
\ArrowLine(70,50)(90,50)
\PhotonArc(50,50)(20,0,180){2}{8}
\Photon(70,50)(70,10){2}{6}
\Vertex(30,50){1.5}
\Vertex(70,50){1.5}
\Text(0,100)[lt]{(b)}
\end{picture}
\begin{picture}(100,100)(0,0)
\ArrowLine(10,50)(30,50)
\ArrowLine(30,50)(70,50)
\ArrowLine(70,50)(90,50)
\PhotonArc(50,50)(20,0,180){2}{8}
\Photon(30,50)(30,10){2}{6}
\Vertex(30,50){1.5}
\Vertex(70,50){1.5}
\Text(0,100)[lt]{(c)}
\end{picture}
\end{center}
\caption{\label{gfig} Corrections to the single gauge field vertex
\eqn{gcoupling}.}
\end{figure}
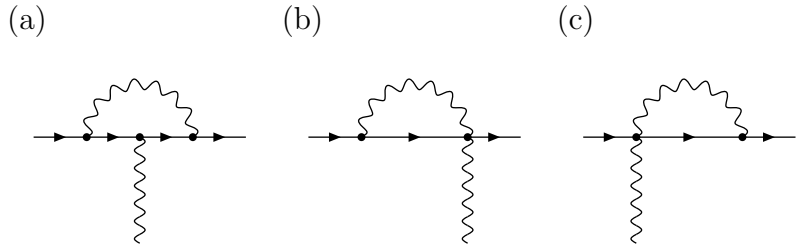
\par
We now consider terms that are dependent on the gauge propagator.  At one
loop, there is just one new graph in the Landau gauge that
could contribute to the scalar propagator (fig.~\ref{prop}):\\
\vbox{\bea\label{gammapsi2}
\Gamma^{(1)}_{\psi^\dagger\psi}
&=&2g^2\int \tripint{k_1}\tripint{k_2} \delta^d({\bf k_1}+{\bf k_2})
\vec\psi^\dagger({\bf k_2})\inner\vec\psi({\bf k_1}) \\
&&\qquad\times\int\tripint{q}(2{\bf k_1} -{\bf q})_i {{i \epsilon_{ijk} q_j}
\over q^2}
(-2{\bf k_2} -{\bf q})_k
{1 \over {\vert{\bf k_1}-{\bf q}\vert^{2-\eta}}} . \nonumber
\eea}
Na\"\i ve power counting with $\eta=0$ would have this graph diverging
logarithmically in three dimensions; however the anomalous dimension
of the scalar field ($\eta \approx 0.016$) makes this graph finite in
the infrared.  In fact, $\Gamma^{(1)}_{\psi^\dagger\psi}$ vanishes
identically because the integrand in \eqn{gammapsi2} is a contraction
of symmetric and antisymmetric tensors.  We conclude that the presence
of the gauge field does not modify the recursion relations for either
$e$ or $\mu$ to one loop.
\par
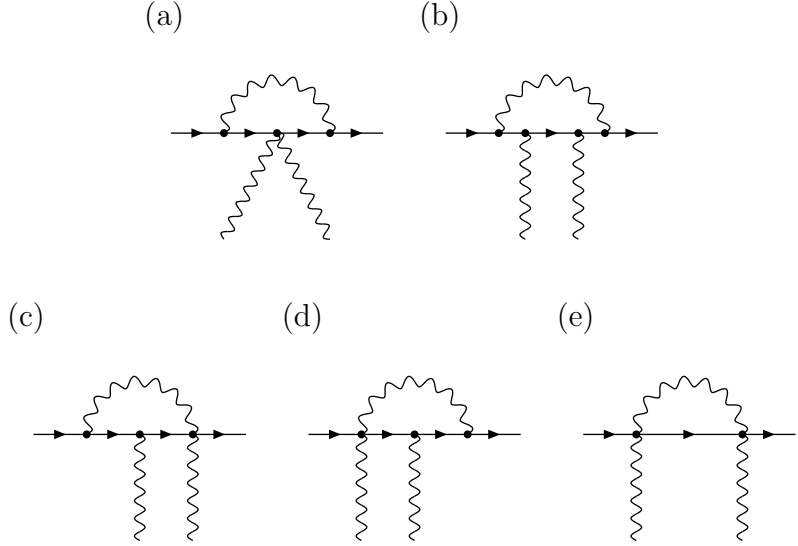
\begin{figure}
\begin{center}
\begin{picture}(100,100)(0,0)
\ArrowLine(10,50)(30,50)
\ArrowLine(30,50)(50,50)
\ArrowLine(50,50)(70,50)
\ArrowLine(70,50)(90,50)
\PhotonArc(50,50)(20,0,180){2}{8}
\Photon(50,50)(30,10){2}{7}
\Photon(50,50)(70,10){2}{7}
\Vertex(30,50){1.5}
\Vertex(50,50){1.5}
\Vertex(70,50){1.5}
\Text(0,100)[lt]{(a)}
\end{picture}
\begin{picture}(100,100)(0,0)
\ArrowLine(10,50)(30,50)
\Line(30,50)(40,50)
\ArrowLine(40,50)(60,50)
\Line(60,50)(70,50)
\ArrowLine(70,50)(90,50)
\PhotonArc(50,50)(20,0,180){2}{8}
\Photon(40,50)(40,10){2}{6}
\Photon(60,50)(60,10){2}{6}
\Vertex(30,50){1.5}
\Vertex(60,50){1.5}
\Vertex(40,50){1.5}
\Vertex(70,50){1.5}
\Text(0,100)[lt]{(b)}
\end{picture}
\end{center}
\begin{center}
\begin{picture}(100,100)(0,0)
\ArrowLine(10,50)(30,50)
\ArrowLine(30,50)(50,50)
\ArrowLine(50,50)(70,50)
\ArrowLine(70,50)(90,50)
\PhotonArc(50,50)(20,0,180){2}{8}
\Photon(70,50)(70,10){2}{6}
\Photon(50,50)(50,10){2}{6}
\Vertex(30,50){1.5}
\Vertex(50,50){1.5}
\Vertex(70,50){1.5}
\Text(0,100)[lt]{(c)}
\end{picture}
\begin{picture}(100,100)(0,0)
\ArrowLine(10,50)(30,50)
\ArrowLine(30,50)(50,50)
\ArrowLine(50,50)(70,50)
\ArrowLine(70,50)(90,50)
\PhotonArc(50,50)(20,0,180){2}{8}
\Photon(30,50)(30,10){2}{6}
\Photon(50,50)(50,10){2}{6}
\Vertex(30,50){1.5}
\Vertex(50,50){1.5}
\Vertex(70,50){1.5}
\Text(0,100)[lt]{(d)}
\end{picture}
\begin{picture}(100,100)(0,0)
\ArrowLine(10,50)(30,50)
\ArrowLine(30,50)(70,50)
\ArrowLine(70,50)(90,50)
\PhotonArc(50,50)(20,0,180){2}{8}
\Photon(30,50)(30,10){2}{6}
\Photon(70,50)(70,10){2}{6}
\Vertex(30,50){1.5}
\Vertex(70,50){1.5}
\Text(0,100)[lt]{(e)}
\end{picture}
\end{center}
\caption{\label{g2fig} Corrections to the quartic gauge vertex
\eqn{g2coupling}.}
\end{figure}
We have already seen that there are no corrections to the gauge field
vertices at one loop order in the absence of the Chern-Simons potential.
We now verify that these vertices remain unchanged in the presence of
the dynamical propagator.  Consider the one loop corrections to
$\Gamma^{(0)}_{A\psi^\dagger\psi}$.  The one loop graphs which
renormalize this vertex are shown in fig.~\ref{gfig}.  As a concrete
example, we consider the first of these (fig.~\ref{gfig}(a)) which
is\\
\vbox{\bea
\Gamma^{(1a)}_{A\psi^\dagger\psi}&=&2g^2
\int\tripint{k_1}\tripint{k_2}\tripint{k_3}
\delta^d({\bf k_1}+{\bf k_2}+{\bf k_3})
A_l({\bf k_3}) \vec\psi^\dagger({\bf k_2})\inner\vec\psi({\bf k_1}) \\
&&\qquad\times\int\tripint{q} ({\bf k_1}+2{\bf q}- {\bf k_2})_l
{{(2{\bf k_1} +{\bf q})_i} \over
{\vert {\bf k_1} - {\bf q} \vert^{2-\eta}}}
{{i \epsilon_{ijk}q_j} \over q^2}
{{(-2{\bf k_2} +{\bf q})_k} \over
{\vert {\bf k_2} + {\bf q} \vert^{2-\eta}}}. \nonumber
\eea}
The relevant part of this graph is the part proportional to ${\bf k_1}
- {\bf k_2}$.  Terms with higher powers of the external momenta
generate irrelevant operators and terms with higher powers of $q$
generate infrared-finite corrections which do not alter the scaling
behavior.  Leaving these uninteresting pieces behind, we now have\\
\vbox{\bea\label{gammaApsi2}
\Gamma^{(1a)}_{A\psi^\dagger\psi}&=&2g^2
\int\tripint{k_1}\tripint{k_2}\tripint{k_3}
\delta^d({\bf k_1}+{\bf k_2}+{\bf k_3})
A_l({\bf k_3}) \vec\psi^\dagger({\bf k_2})\inner\vec\psi({\bf k_1}) \\
&&\qquad\times\int\tripint{q}({\bf k_1}-{\bf k_2})_l  {{i \epsilon_{ijk} q_i
q_j q_k}
\over q^{6-2\eta}}. \nonumber
\eea}
Once again, a non-zero $\eta$ saves us from having a potentially
logarithmic infrared divergence for $d=3$.  But again, the relevant
part of $\Gamma^{(1a)}_{A\psi^\dagger\psi}$ vanishes due to the
antisymmetry of the Chern-Simons propagator.  By power counting, the
other two graphs (figs.~\ref{gfig}(b) and
\ref{gfig}(b)) can be shown to generate (at worst) finite corrections.
These graphs in fact vanish as they are both integrals of odd
functions.
\par
\begin{figure}
\begin{center}
\begin{picture}(100,100)(0,0)
\Text(0,100)[lt]{(a)}
\ArrowLine(10,70)(30,70)
\ArrowLine(30,70)(70,70)
\ArrowLine(70,70)(90,70)
\ArrowLine(10,30)(30,30)
\ArrowLine(30,30)(70,30)
\ArrowLine(70,30)(90,30)
\Vertex(30,70){1.5}
\Vertex(70,70){1.5}
\Vertex(30,30){1.5}
\Vertex(70,30){1.5}
\DashLine(30,70)(30,30){1.5}
\Photon(70,70)(70,30){2}{6}
\end{picture}
\begin{picture}(100,100)(0,0)
\Text(0,100)[lt]{(b)}
\ArrowLine(10,70)(30,70)
\ArrowLine(30,70)(70,70)
\ArrowLine(70,70)(90,70)
\ArrowLine(10,30)(30,30)
\ArrowLine(30,30)(70,30)
\ArrowLine(70,30)(90,30)
\Vertex(30,70){1.5}
\Vertex(70,70){1.5}
\Vertex(30,30){1.5}
\Vertex(70,30){1.5}
\Photon(30,70)(30,30){2}{6}
\DashLine(70,70)(70,30){1.5}
\end{picture}
\begin{picture}(100,100)(0,0)
\Text(0,100)[lt]{(c)}
\ArrowLine(10,70)(30,70)
\ArrowLine(30,70)(70,70)
\ArrowLine(70,70)(90,70)
\ArrowLine(90,30)(70,30)
\ArrowLine(70,30)(30,30)
\ArrowLine(30,30)(10,30)
\Vertex(30,70){1.5}
\Vertex(70,70){1.5}
\Vertex(30,30){1.5}
\Vertex(70,30){1.5}
\Photon(30,70)(30,30){2}{6}
\DashLine(70,70)(70,30){1.5}
\end{picture}
\end{center}
\begin{center}
\begin{picture}(100,100)(0,0)
\Text(0,100)[lt]{(e)}
\ArrowLine(10,70)(30,70)
\ArrowLine(30,70)(70,70)
\ArrowLine(70,70)(90,70)
\ArrowLine(10,30)(30,30)
\ArrowLine(30,30)(70,30)
\ArrowLine(70,30)(90,30)
\Vertex(30,70){1.5}
\Vertex(70,70){1.5}
\Vertex(30,30){1.5}
\Vertex(70,30){1.5}
\Photon(30,70)(30,30){2}{6}
\Photon(70,70)(70,30){2}{6}
\end{picture}
\begin{picture}(100,100)(0,0)
\Text(0,100)[lt]{(f)}
\ArrowLine(10,70)(30,70)
\ArrowLine(30,70)(70,70)
\ArrowLine(70,70)(90,70)
\ArrowLine(90,30)(70,30)
\ArrowLine(70,30)(30,30)
\ArrowLine(30,30)(10,30)
\Vertex(30,70){1.5}
\Vertex(70,70){1.5}
\Vertex(30,30){1.5}
\Vertex(70,30){1.5}
\Photon(30,70)(30,30){2}{6}
\Photon(70,70)(70,30){2}{6}
\end{picture}
\begin{picture}(100,100)(0,0)
\Text(0,100)[lt]{(d)}
\ArrowLine(10,70)(30,70)
\ArrowLine(30,70)(50,70)
\ArrowLine(50,70)(70,70)
\ArrowLine(70,70)(90,70)
\PhotonArc(50,70)(20,0,180){2}{8}
\ArrowLine(10,30)(50,30)
\ArrowLine(50,30)(90,30)
\DashLine(50,30)(50,70){1.5}
\Vertex(50,30){1.5}
\Vertex(30,70){1.5}
\Vertex(50,70){1.5}
\Vertex(70,70){1.5}
\end{picture}
\begin{picture}(100,100)(0,0)
\Text(0,100)[lt]{(g)}
\ArrowLine(10,70)(50,70)
\ArrowLine(50,70)(90,70)
\ArrowLine(10,30)(30,30)
\ArrowLine(30,30)(70,30)
\ArrowLine(70,30)(90,30)
\Vertex(50,70){1.5}
\Vertex(30,30){1.5}
\Vertex(70,30){1.5}
\Photon(50,70)(30,30){2}{7}
\Photon(50,70)(70,30){2}{7}
\end{picture}
\end{center}
\caption{\label{ufig} Vanishing corrections to the $u$ vertex.}
\end{figure}
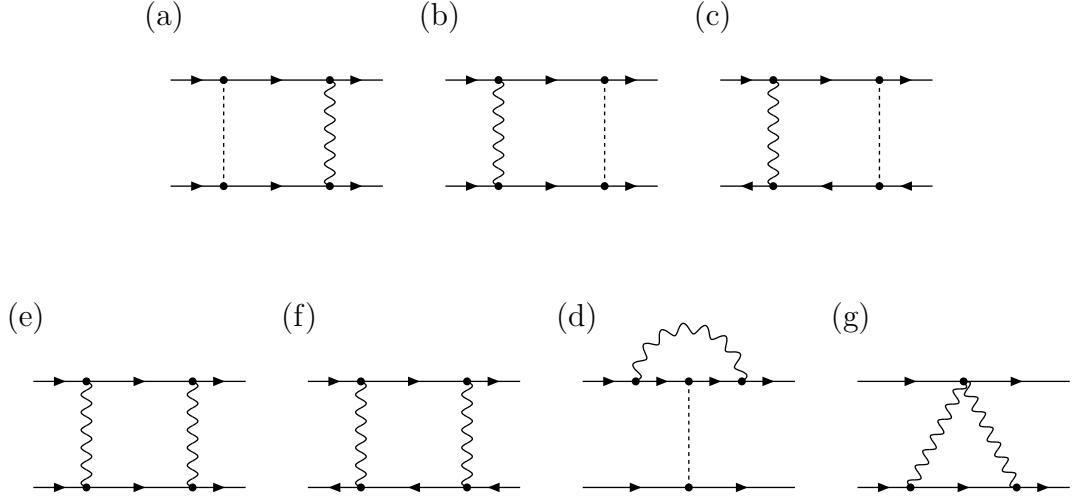
The possible corrections to the quartic gauge coupling \eqn{g2coupling}
are depicted in fig.~\ref{g2fig}.  None of these graphs will contribute to
the renormalization of the vertex however:  by power counting arguments
they are all finite at the Wilson-Fisher fixed point.  Indeed
the relevant parts of all of these graphs
vanish in Landau gauge due to the antisymmetric gauge propagator.
\par
\begin{figure}[t]
\begin{center}
\begin{picture}(100,100)(0,0)
\ArrowLine(10,30)(50,30)
\ArrowLine(50,30)(90,30)
\ArrowLine(10,70)(50,70)
\ArrowLine(50,70)(90,70)
\Vertex(50,70){1.5}
\Vertex(50,30){1.5}
\PhotonArc(50,50)(20,90,270){2}{8}
\PhotonArc(50,50)(20,270,90){2}{8}
\end{picture}
\end{center}
\caption{\label{ufig2} A finite one loop correction to the $u$ vertex.}
\end{figure}
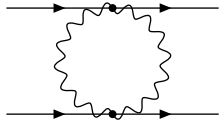
The one loop results just obtained are natural consequences of gauge
invariance.  We must now consider the $u$ vertex which is marginally
relevant at the Wilson-Fisher fixed point.  We will verify that the
fixed point value for $u$ does not lose its stability with the
introduction of the gauge propagator.  To see this, we consider the
one loop graphs depicted in figs.~\ref{ufig} and \ref{ufig2}.  Once
again power counting tells us that none of these graphs can generate a
relevant correction to scaling ({\sl i.e.} an infrared-divergent
graph).  Moreover, the relevant parts of all the graphs in
fig.~\ref{ufig} vanish, once again due to the antisymmetry of the
gauge propagator.  The graph in fig.~\ref{ufig2} on the other hand,
does not vanish and generates a finite correction to $u$:
\vbox{\bea\label{finiteloop}
u&=&u_0 +4g^4 \int\tripint{q}
{{\epsilon_{kji} \epsilon_{ilk} q_j q_l} \over q^4} \nonumber \\
&=&u_0-4g^4 \int_0^\Lambda {{q^{d-1}dq}\over{(2\pi)^d}}
{{(d-1)f(d)}\over q^2} \nonumber \\
&\stackrel{d=3}{\longrightarrow}&u_0- {4{g^4} \over {\pi^2 L_{\rm p}}} .
\eea}
Here $f(d)$ is the area of a unit $d$-sphere and the persistence length of the
chain $L_{\rm p}$ defines the short-distance, ultraviolet cutoff
$\Lambda=L_{\rm p}^{-1}$.  This term decreases the amount of
self-avoidance in our random walk by a finite amount.  Since a finite
correction does not change the recursion relations in \eqn{recursion},
this correction  will not change the scaling exponent for the
radius of gyration.  Nonetheless, its presence can be felt if
the chemical potential for writhe $g^2$ exceeds a certain
threshold.  When $4g^4 L_{\rm p}^{-1}\pi^{-2} \geq u_0$ (where $u_0$
is the bare repulsion) the repulsive term
becomes attractive
and the chain collapses as it would in a bad solvent.
In the case of excluded volume interactions $u_0
\sim L_{\rm p}$ ({\sl i.e.} $k_B T$ per persistence length) and the chain
will collapse for $g^4\grsim (\pi/2)^2$.
\par
We have thus shown that the presence of a Chern-Simons field does not
change the critical behavior of the $O(N)$ complex scalar model in the
$N \rightarrow 0$ limit at one loop.  Using the mappings of
sections 2 and 3, we can interpret this results in
terms of closed self-avoiding walks with a chemical potential for
link.  The presence of such a chemical potential
does not affect the statistics of the random walk unless it
exceeds a certain threshold, at which point the chain collapses.

\section{Average Writhe of a Closed Loop}

\par We have found that the presence of a small chemical potential for
writhe does not change the scaling behavior of a self-avoiding
random walk.  Here we study the first two moments of the writhe,
$\langle \Wr \rangle$ and $\langle \Wr^2 \rangle$.
\par
In the absence of a chemical potential $g^2$, a rigorous result
establishes that the ensemble average of the magnitude of the writhe
must scale at least as fast as the square root of the chain length
\cite{Rensburg1993}.  Numerical results saturate this bound: $\langle
\vert \Wr \vert \rangle \sim L^{1 \over 2}$ \cite{Orlandini1994}.
We therefore expect the average of the writhe squared to scale with
the length of the polymer.  A simple estimate of $\langle \Wr^2
\rangle$ can be obtained by assuming that the tangents are perfectly
correlated within a persistence length and completely uncorrelated at
larger distances \cite{Marko.disc}.  We have
\vbox{
\bea\label{writhe2}
\langle \Wr^2 \rangle&=&{1 \over 16\pi^2Z}\int [d{\bf R}]
e^{-F} \int ds_1 ds_2 ds_3 ds_4
\left({{d{\bf R}(s_1)} \over {ds_1}} \cross
{{d{\bf R}(s_2)} \over {ds_2}} \right) \inner {{{\bf R}(s_1)-{\bf R}(s_2)}
\over {\vert {\bf R}(s_1)-{\bf R}(s_2)\vert^3}} \nonumber \\
&&\qquad\times\left({{d{\bf R}(s_3)} \over {ds_3}} \cross
{{d{\bf R}(s_4)} \over {ds_4}} \right) \inner
{{{\bf R}(s_3)-{\bf R}(s_4)} \over {\vert {\bf R}(s_3)-{\bf R}(s_4)\vert^3}}
\eea}
where the free energy cost is just a bending stiffness:
\beq
F[{\bf R}]={\kappa\over 2}\int ds\left( {d^2{\bf R}} \over {ds^2} \right) .
\eeq
Our simplifying assumption is
\beq
\left\langle{{{dR_i(s)} \over {ds}}\, {{dR_j(s')} \over {ds'}}}\right\rangle =
\left\{\begin{array}{r@{\quad}l}
\delta_{ij} & {\rm for}\,\vert s-s'\vert<L_{\rm p} \\
0 & {\rm otherwise}
\end{array}\right. .
\eeq
Using this tangent-tangent correlation function and neglecting the
correlations between ${\bf R}$ and $d{\bf R}/ds$, we find
\bea
\langle \Wr^2\rangle
&\sim& \int ds_1 ds_2 ds_3 ds_4
\left\langle{{({\bf R}(s_1)-{\bf R}(s_2))} \over
{\vert {\bf R}(s_1)-{\bf R}(s_2)\vert^3}} \inner
{{({\bf R}(s_3)-{\bf R}(s_4))} \over
{\vert {\bf R}(s_3)-{\bf R}(s_4)\vert^3}}\right\rangle \nonumber \\
&\sim&L_{\rm p}^2 \int ds_1 ds_2 {1\over{\vert{\bf R}(s_1)-{\bf
R}(s_2)\vert^4}} \nonumber \\
&\sim&L_{\rm p}^2 \int ds_1 ds_2 {1\over{L_{\rm p}^2 \vert s_1
-s_2\vert^2}}\nonumber \\
&\sim&L\left({1\over L_P}- {1\over L}\right) \sim  {L\over{L_{\rm p}}} .
\eea
Here we have used the statistics for a Gaussian chain in the last
step: $\langle \vert{\bf R}(s)-{\bf R}(s')\vert^2\rangle \sim L_{\rm
p}\vert s-s'\vert$.  Using the self-avoiding scaling behavior
$\langle \vert{\bf R}(s)-{\bf R}(s')\vert^2\rangle \sim L_{\rm
p}^{2-2\nu}\vert s-s'\vert^{2\nu}$ gives precisely the same result.
We thus see that short-distance correlations saturate the rigorous lower
bound.
\par
Our long wavelength theory makes a different prediction for the
scaling behavior of $\langle \Wr^2 \rangle$.  In section \ref{mapping},
we found the correlation function $G^C({\bf x};\mu)$ for a closed
polymer loop with a chemical potential $\mu$ per monomer
\eqn{ewillii}.  Using this function as a starting point, we
determine the scaling behavior of the moments of the writhe.  After
taking the continuum limit for large $L$, we recover the fixed
length generating function using an inverse Laplace transform:
\beq\label{upsilon}
\overline{\Upsilon}^C(L)={1 \over {2 \pi i}} \oint
\overline{G}^C(\mu;g^2) {\rm e}^{-\mu (L-1)} d\mu
\eeq
where the contour of integration encloses all the poles of the integrand and
\beq
\overline{G}^C(\mu;g^2)=\lim_{N \rightarrow 0} {1 \over N}
\left\langle{\vec\psi^\dagger({\bf 0})\inner\vec\psi({\bf 0})}\right\rangle
\eeq
is the generating function for all closed loops.  From this function,
the second moment of the writhe for the closed paths is easily
obtained:
\beq\label{avgwrithe2}
\langle \Wr^2 \rangle=\left.   {{d^2} \over {d(g^2)^2}}
\ln\,\overline{\Upsilon}^C({\bf x}; L) \right\vert_{g^2=0}.
\eeq
Explicit evaluation of the integral in \eqn{upsilon} is difficult, but
fortunately we are only interested in the way $\Upsilon^C$ scales with
$L$.  We assume a scaling relation for the correlation function,
\beq
\overline{G}^C(\mu;g^2) = L^{\alpha-1} \overline{G}^C(\mu L;g^2),
\eeq
where $\alpha$ is the specific heat exponent and we have used the fact
that $g^2$ does not rescale.  We can now obtain the scaling behavior of
$\overline{\Upsilon}^C$:
\begin{figure}
\centerline{\psfig{figure=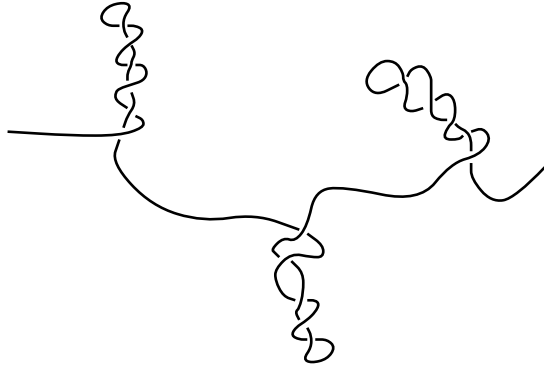,width=3.5in}}
\caption{The writhe is dominated by short
plectonemic regions separated by distances larger than the persistence length.}
\label{plectonemes}
\end{figure}
\bea
\overline{\Upsilon}^C(L)&=&{1 \over{2\pi i}} \oint L^{\alpha-1}
\overline{G}^C(\mu L;g^2) {\rm e}^{-\mu L}
{{d(\mu L)} \over L} \nonumber \\
&\sim&L^{\alpha-2} .
\eea
As we saw in the last section, the chemical potential for writhe
$g^2$ does not get renormalized.  What is more, this chemical
potential does not appear at this order in the scaling exponent of
$\overline{\Upsilon}^C$.  It follows that all of the $L$ dependence
drops out of \eqn{avgwrithe2} when the derivatives of the logarithm
are taken.  We conclude that $\langle \Wr^2\rangle$ does not scale
linearly with the length of the chain; at best, the length-dependence
enters as some power of a logarithm.  Another paradoxical result is
obtained when we compute the average writhe in the presence of a
chemical potential:
\beq
\langle \Wr \rangle = - {d\over{d(g^2)}}\ln\,\Upsilon^C(L)\sim L^0
\eeq
The length dependence drops out just as it did for the second moment.
A chemical potential for writhe does not seem to require the total
amount of writhe to grow with the chain length.
\par
These results, though at first surprising, are consistent with
existing results.  By construction, our field theory only applies to
the behavior of the random walk on scales longer than the persistence
length.  We have seen that the writhe is concentrated on shorter
length scales, and so we would not expect to see it in this field
theoretic treatment.  A picture emerges of short plectonemic regions
separated from each other by at least the polymer persistence length
(see fig.~\ref{plectonemes}).  These plectonemic regions each
contribute to the writhe of the polymer, but these contributions
appear only as matching conditions in our long wavelength theory.  We
would therefore expect that in the presence of a chemical potential
for writhe, the average writhe of the chain {\sl will} scale with the
chain length:
\beq\label{speculation}
\langle \Wr \rangle \approx {L \over {L_{\rm p}}} \langle \Wr \rangle_{\rm pl},
\eeq
where $\langle \Wr \rangle_{\rm pl}$ is the average writhe per
plectonemic segment.  Likewise, the second moment is dominated by the
small plectonemic regions:
\beq
\langle \Wr^2\rangle \approx {L\over L_{\rm p}}\langle \Wr^2 \rangle_{\rm pl}.
\eeq
It might seem strange to talk about the writhe of a short segment of
the polymer when writhe is a global property of the entire chain.  But
as can be seen from the $\vert {\bf R}(s)-{\bf R}(s')\vert^{-3}$ factor in
\eqn{ewri}, the writhe is dominated by the correlations between
regions that are close together in space.

\section{Conclusion}

\par In this paper we have investigated the topology of
self-avoiding random walks.  Our goal was to understand how the
linking number constraint changes the conformational statistics of
closed polymers.  After integrating out the twist degree of freedom,
this problem was shown to be equivalent to that of studying the
statistics of a polymer with a chemical potential for writhe.  By
mapping this modified problem onto the $N \rightarrow 0$ limit of an
$O(N)$ complex scalar field theory with a Chern-Simons gauge field, we
were able to find that the presence of a small chemical potential for
writhe does not change the behavior of our polymer in the scaling
limit: that is, the renormalization group equations for the ungauged
scalar field model that describes the usual self-avoiding walk do not
get modified by the introduction of the Chern-Simons term.  We also
saw that a sufficiently large chemical potential, $g^2 \grsim (\pi/2)$,
collapses the polymer.
\par
We carried out our perturbation expansion to one loop.  The anomalous
wavefunction renormalization of $\psi$ at the Wilson-Fisher fixed
point kept all of the possible corrections to scaling finite; this
result should be compared to that of na\"\i ve power counting where
logarithmic divergences occur.  In view of this, it is reasonable to
believe that the renormalization group equations do not change to all
orders in the loop expansion.  It would be interesting to carry out a
careful analysis of the power counting at the Wilson-Fisher fixed
point.  It may also be possible to adopt the wealth of knowledge on anyons
\cite{Pryadko} to study the critical behavior of our model as well as
the use of supersymmetry \cite{Parisi} to study self-avoiding walks.
\par
Additionally, we investigated the scaling behavior of the writhe itself.  We
found that the long wavelength contribution to the writhe does not
scale with its length.  The writhe escapes the notice of the scaling
theory by hiding at small scales.  We find our results to be
consistent with a picture whereby the writhe is confined mainly to
short plectonemic segments separated by at least the bend persistence length.
\par
Our overall results indicate that the effects of topology do not
change the general scaling properties of polymers: constrained random
walks belong to the same universality class as unconstrained ones.
The effects of topology seem to enter only on small scales through
the chemical and physical details of the chain.
\par
Recently \cite{Nelson} the winding number statistics of directed
polymers have been analyzed by adding an auxiliary $2+1$ dimensional
gauge field.
We note that the Chern-Simons formalism presented here can be
used to study the linking number statistics of an ensemble of closed,
{\sl isotropic},
polymer loops, as can be found, for instance, in the kinetoplast DNA
of trypanasomes \cite{Kreuzer}.  The rheological properties of these
so-called ``olympic gels'' are intimately connected to their topology.
Thus, via bulk experiments, the linking number statistics could be
explored.

\section{Acknowledgments}
It is a pleasure to acknowledge stimulating discussions with M.~Goulian,
G.~Jungman,
T.C.~Lubensky, J.~Marko, D.R.~Nelson, P.~Nelson, C.S.~O'Hern, T.~Powers
and J.~Toner.
JDM was
supported by NSF Grant DMR95-07366 and an FCAR graduate fellowship from
the government of Qu\'{e}bec.
RDK thanks The Rockefeller University Center For
Studies In Physics and Biology, where some of this work was done.  RDK
was supported by NSF Grants DMR94-23114 and DMR91-22645.

\bigskip

\newcommand{\noopsort}[1]{} \newcommand{\printfirst}[2]{#1}
  \newcommand{\singleletter}[1]{#1} \newcommand{\switchargs}[2]{#2#1}

\end{document}